\documentclass[11pt]{article}
\usepackage[T1]{fontenc}
\usepackage[a4paper]{geometry}
\usepackage{amssymb}
\usepackage[normalsize,normal,sc]{caption}

\usepackage[latin1]{inputenc}

\def \build#1#2#3{\mathrel{\mathop{\kern 0pt#1}\limits_{#2}^{#3}}}

\def \j{j^\star}
\def \no{\noindent}
\def \dis{\displaystyle}
\def \l{\left}
\def \r{\right}

\def \beq{\begin{equation}}
\def \eeq{\end{equation}}
\def \be{\begin{eqnarray*}}
\def \ee{\end{eqnarray*}}
\def \ben{\begin{eqnarray}}
\def \een{\end{eqnarray}}
\def \pl{\medskip}

\def \B{\Big}
\def \b{\big}
\def\qE{\mathbb{E}}
\def\qe{\varepsilon}

\def\qZ{\mathbb{Z}}

\def\qP{\mathbb{P}}

\def\bibfmta#1#2#3#4{{#1}, {#2}, {\em #3}, #4.}
\def\bibfmtb#1#2#3#4{{#1}, {\em #2}, {#3}, #4.}

\begin{document}

\newtheorem{fig}{\hspace{2cm} Figure}
\newtheorem{lem}{Lemma}
\newtheorem{defi}{Definition}
\newtheorem{pro}{Proposition}
\newtheorem{rem}{Remark}
\newtheorem{theo}{Theorem}
\newtheorem{cor}{Corollary}
\newenvironment{conj}{\newline\bf Conjecture:\sl}{\rm}
\newenvironment{proof}{\no \textbf{Proof:} }{\hfill $\square$}
\newcounter{a}[subsection]

\begin{center}
\LARGE{Quasi-Optimal Leader Election Algorithms in Radio Networks \\
with Log-logarithmic Awake Time Slots}
\pl

\bigskip \large Christian \textsc{Lavault}\footnote{LIPN %
(CNRS UMR 7030), Universit\'e Paris~13, 99, av. J.-B. Cl\'ement 93430
Villetaneuse France.\ E-mail: \{lavault,vlad\}@lipn.univ-paris13.fr}, %
Jean-Fran\c{c}ois \textsc{Marckert}\footnote{LAMA (CNRS UMR 8100), %
Universit\'e de Versailles St-Quentin en Yvelines, %
45, av. des Etats-Unis 78035 Versailles France.\ E-mail: marckert@math.uvsq.fr}, %
Vlady \textsc{Ravelomanana}$^1$

\medskip
{\small (January 1st, 2003)}
\end{center}

\bigskip
\begin{quotation}
\small \noindent \bf ABSTRACT:\ \rm A radio network (RN) is a
distributed system consisting of $n$ radio stations. We  design
and analyze two distributed leader election protocols in RN where
the number $n$ of  radio stations is unknown. The first algorithm
runs under the assumption of {\it limited collision detection},
while the second assumes that {\it no collision detection} is
available. By ``limited collision detection'', we mean that if
exactly one station sends (broadcasts) a message, then all
stations (including the transmitter) that are listening at this
moment receive the sent message. By contrast, the second
no-collision-detection algorithm assumes that a station cannot
simultaneously send and listen signals. Moreover,  both protocols
allow the stations to keep asleep as long as possible, thus
minimizing their awake time slots (such algorithms are called {\it
energy-efficient}). Both randomized protocols in RN are shown to
elect a leader in  $O(\log{(n)})$ expected time, with no station
being awake for more than $O(\log{\log{(n)}})$ time slots.
Therefore, a new class of efficient algorithms is set up that
match the $\Omega(\log{(n)})$ time lower-bound established by
Kushilevitz and Mansour in~\cite{Mansour}.
\normalsize
\end{quotation}

\section{Introduction}
Electing a leader is a fundamental problem in distributed systems
and it is studied in a variety of contexts including radio
networks~\cite{CHLEBUS}. A radio network (RN, for short) can be
viewed as a distributed system of $n$ radio stations with no central
controller. The stations are bulk-produced, hand-held devices and
are also assumed to be indistinguishable: no identification numbers
(or IDs) are available. A large body of research has already focused
on finding efficient solutions to elect one station among an
$n$-station RN under various assumptions (see
e.g.~\cite{CHLEBUS,Mansour,Willard}). It is also assumed that the
stations run on batteries. Therefore, saving battery power is important,
since recharging batteries may not be possible in standard working conditions.
We are interested in designing \textit{power-saving} protocols (also called
\textit{energy-efficient} protocols). The present work is motivated by
various applications in emerging technologies: from wireless communications,
cellular telephony, cellular data, etc., to simple hand-held multimedia
services~\cite{BLACK}.

\pl \no
\textbf{The models.}\ As customary, time is assumed to be slotted,
stations work synchronously and have no IDs available. No \textit{a priori}
knowledge is assumed on the number $n \ge 2$ of stations involved in the
RN: neither a (non-trivial) lower-bound nor an upper-bound on $n$.
Awake stations areallowed to communicate globally (i.e.~the underlying graph
is a clique) by using a unique radio frequency channel with no collision
detection (no-CD for short) mechanism. If, during a step, stations may
either send (broadcast) a message or listen to the channel, then we talk
about {\it weak} no-CD RN model.  If both operations can be performed
simultaneously, then the model is called the {\it strong} no-CD RN. Namely,
if exactly one station sends, then all stations that listen at this time slot,
including the transmitter, receive the message. (In the literature, no-CD
RN usually means strong model, see e.g.~\cite{Mansour,OLARIU}.)
Such models feature concrete situations; in particular, the lack of feedback
mechanism experiences real-life applications (see e.g.~\cite{NOCOLLISION}).
Usually, the natural noise existing within radio channels makes it
impossible to carry out message collision detection. It is thus highly
desirable to design protocols that do not depend on the reliability of any
collision detection mechanism. When sleeping, any given station remains
unable to hear another station, and it may also keep unaware of the election
instant time in the protocol. However, stations (awake or asleep) are
{\em all} required to become eventually aware of the final status of the RN.
More precisely, each station may be in two states:

$\bullet$\ either \textit{awake}, i.e. listening and/or broadcasting,
according to the respective model (weak or strong no-CD RN),

$\bullet$\ or \textit{asleep}, and thus saving its own battery. When
sleeping, a station is ``out of reach'': it cannot be waked up by none
of its neighbours.

Note also that each broadcast finishes within a rather short lapse of time,
and that each awake receiver is able to check if a signal has been sent
by exactly one station.

\pl \no \textbf{Related works.}\ The RN model considered herein
may be regarded as a broadcast network model (see
e.g.~\cite{CHLEBUS}). In this setting, e.g.,
Willard's~\cite{Willard}, Greenberg's \textit{et
al.}~\cite{PHILIPPE-ACM} (with collision detection) and
Kushilevitz and Mansour~\cite{Mansour} (no-CD) are among the most
popular leader elections protocols. In the model,
\cite{NOCOLLISION} may serve as a global reference for basic
conflict-resolution based protocols. Previous researches on
multiple-access channel mainly concern stations that are kept
awake during the whole of a protocol in the RN, even when such
stations are the ``very first losers'' of a coin flipping game
algorithm~\cite{Helmut}. In~\cite{Polonais}, the authors design an
energy-efficient protocol (with $o\l(\log\log(n)\r)$ energy cost)
that approximate $n$ up to a constant factor, but with running
time $O\l(\log^{2+\epsilon}(n)\r)$ in strong no-CD RN. Also,
distributional analyses of various randomized election protocols
may include~\cite{Szpankowski-Fill-Mahmoud,Szpankowski-Janson} for
example.

\pl \no
\textbf{Our results.}\ The first leader election protocol (Algorithm~1)
presented in the paper runs in the {\em strong} no-CD RN model, while
the second one (Algorithm~2) works in the {\em weak} no-CD RN model.
We design a class of double-loop leader election algorithms that achieve
an average $O(\log{n})$ running time complexity and an average
$O(\log{\log{n}})$ awake time slots for each station in the RN.
Indeed, both algorithms match the $\Omega(\log{n})$ time lower-bound
established in~\cite{Mansour} and also allow the stations to keep
sleeping most of the time. In other words, each algorithm greatly reduces
the total awake time slots of the $n$ stations: shrinking from the usual
$O(n\log{n})$ downto $O(n\log\log{n})$, while their expected time
complexity still is $O(\log{n})$ (with respect to the execution
time). Our protocols are thus ``energy-efficient'' and suitable for
hand-held devices working with batteries.
Besides, the algorithms use a parameter $\alpha$ which works as a
precise and flexible regulator. By tuning the value of $\alpha$,
the running time ratio of each protocol to its energy consumption
may be adjusted ($\alpha$ serves a ``potentiometer).
Furthermore, the design of Algorithms~1 and 2 suggests that within
both weak and strong no-CD RN, the mean time complexity of the
algorithms only differs of a constant factor. Also, our results
improve on~\cite{OLARIUCONF}.

\bigskip \no
\textbf{Outline of the paper.}\ In Section~2, we first present Algorithms 1
and 2, which use a simple coin-tossing procedure ({\em rejection algorithms}).
Section~3 is devoted to the analyses of both algorithms, by means of tight
asymptotics techniques. We conclude in Section~4.

\section{Algorithms and Results}
Both algorithms rely  on the intuitive evidence that each station must
be awake within a sequence of predetermined time slots. A first naive idea is
to have stations using probabilities $1/2$, $1/4$,\ldots to wake up and broadcast.
This solution is not correct however, since it is possible that no station ever
broadcasts alone.

In order to correct the failure, we have to plan many rounds with
predetermined length. Awake time slots are programmed at the end of
each such rounds. Thus, we allow all stations to detect the (possible)
termination of the session in each round.
In the sequel, we let $\alpha>1$ be the tuning parameter.

\subsection{Algorithm 1}

\pl \no
\hrule
\vspace{0.05cm} \no
( 1)\indent $round \leftarrow 1$; \\
( 2)\indent {\bf Repeat} \\
( 3)\indent \indent {\bf For}\ $k$ from 1 to $\lceil \alpha^{round} \rceil$\ %
            {\bf do} \hfill /* probabilistic phase */ \\
( 4)\indent \indent \indent Each station wakes up independently with %
            probability $1/2^{k}$ (to broadcast {\em and} listen); \\
( 5)\indent \indent \indent \textbf{If}\ a unique station broadcasts\ {\bf then}\ %
            it becomes a {\sl candidate} station {\bf EndIf}; \\
( 6)\indent \indent {\bf EndFor} \hfill /* deterministic phase */ \\
( 7)\indent \indent At the end of each round, all stations wake up, %
                listen and all {\sl candidate} stations broadcast; \\
( 8)\indent \indent {\bf If}\ there is a unique {\sl candidate}\ {\bf then}\ %
                it is {\em elected}\ {\bf EndIf}; \\
( 9)\indent \indent $round \leftarrow round + 1$; \\
(10)\indent {\bf until a station is elected} \vspace{0.25cm} \hrule
\begin{center}
\textbf{Algorithm 1. Leader election protocol for strong no-CD RN}
\end{center}
\no
Given a round $j$ in the outer-loop (repeat-until loop), during the
execution time of the inner-loop each station randomly chooses to sleep
or to broadcast (and/or to listen) at each time slot (each station can
compute its sequence of awaking times at the beginning of a current round).
If a unique station is broadcasting, this station knows the status of
the radio channel and it becomes a {\sl candidate}. At the end of round
$j$, every station wakes up and listens to the channel; then the candidates
broadcast. If there is a single candidate, it is elected. Otherwise,
the next round begins.

Define $q$ as the probability of having an election after
$j^{\star}(n) = \lceil \log_{\alpha}{\log_2{n}} \rceil$ rounds and let $c_q$
be the function defined in inequalities~(\ref{doublesum}) and~(\ref{ESPERANCE}),
\begin{equation} \label{CC}
c_q(\alpha) = \frac{q {\alpha}^3}{(\alpha-1)\b(1-\alpha (1-q)\b)}\,.
\end{equation}
\begin{theo} \label{Theorem-ALGO1}
On the average, Algorithm 1 elects a leader in at most $c(\alpha,q) \log_2{n}$
time slots, with no station being awake for more than
$2 \log_{\alpha} \log_{2}(n)\,\b(1+o(1)\b)$ mean time slots,
where $c_{q_1}(\alpha)$ is given in~(\ref{CC}) with $q_1 = .6305$.
\end{theo}

\subsection{Algorithm 2}
In the case of weak no-CD RN, a potential candidate cannot alone be
aware of its status since it cannot broadcast {\em and} listen at the same time.
So, {\sl witnesses} are needed to inform the candidates.
\pl \no
\hrule
\vspace{0.05cm} \no
( 1)\indent $round \leftarrow 1$; \\
( 2)\indent {\bf Repeat}\\
( 3)\indent \indent {\bf For}\ $k$ from 1 to $\lceil \alpha^{round} \rceil$\ {\bf do} %
                \hfill /* probabilistic phase */ \\
( 4)\indent \indent \indent  Each station wakes up independently %
                with probability $1/2^k$; \\
( 5)\indent \indent \indent With probability $1/2$ each awake station decides \\
( 6)\indent \indent \indent {\em either} to broadcast the message %
            $\langle ok\rangle$ {\em or} to listen; \\
( 7)\indent \indent \indent A listening station that gets this message %
            (from one single sender) becomes a {\sl witness}; \\
( 8)\indent \indent {\bf EndFor} \hfill /* deterministic phase */ \\
( 9)\indent \indent  At time $\lceil \alpha^{round} \rceil+1$, each witness %
         and each station having broadcasted wakes up; \\
(10)\indent \indent Each witness broadcasts (forwards) its received message; \\
(11)\indent \indent If there is one single witness, the station that sent %
        the (``witness'') message $\langle ok\rangle$ is elected; \\
(12)\indent \indent  At time $\lceil \alpha^{round} \rceil+2$, %
             all stations are listening; \\
(13)\indent \indent \indent  \textbf{If}\ the leader has been elected\ %
            \textbf{then}\ the leader broadcasts \\
(14)\indent \indent \indent  and all stations are aware of the status\ \textbf{EndIf}; \\
(15)\indent \indent $round \leftarrow round + 1$; \\
(16)\indent {\bf until a station is elected.}
\vspace{0.25cm}
\hrule

\begin{center} \textbf{Algorithm 2. Leader election protocol for weak no-CD RN}
\end{center}
\no
This algorithm is in the same vein as Algorithm~1. Yet, in Algorithm~2
no candidate can listen to its own message. Therefore, to be elected,
a candidate needs the help of a witness. It is important to remark that,
in line~(7), a station is defined as a {\it witness} {\em iff} it wakes
up {\em exactly} when there exists a {\em single} broadcasting station.
The election thus takes place at the end of the round during which
two stations are chosen among $n$, viz. the single candidate and its
corresponding witness.

Some modifications in Algorithm~2 would slightly improve its performances.
For example, to avoid possible conflicts, witnesses could be kept asleep
till the end of each round and also, the algorithm could prevent any
broadcasting station from becoming a witness. In its present form,
we have the following result.

\begin{theo} \label{Theorem-ALGO2}
On the average, Algorithm~2 elects a leader in at most $c_{q_2}(\alpha)\log_2(n)$
time slots, with no station being awake for more than
$2.5 \log_{\alpha} \log_{2}(n)\,\b(1+o(1)\b)$ mean time slots,
where $c_{q_2}(\alpha)$ is given by~(\ref{CC}), with $q_2 = .6176$.
\end{theo}

\section{Analysis}
\subsection{Technical Lemmas} \label{TECHLEM}
The following two Lemmas use Mellin transforms~\cite{PHILIPPE-ROBERT,KNUTH};
they are both at the basis of our analyses.
\begin{lem} \label{LEMME-MELLIN1}
We have
\ben
\sum_{k=1}^{r} \frac{n}{2^k}\exp{\l(-\frac{n}{2^k}\r)} = \frac{1}{\log{2}} %
\;+\;\frac{1}{\log{2}}\,U(\log_2{n})\;+\; O\l(\frac{n}{2^r}\r)\;+\; O\l(\frac{1}{n}\r),
\label{EQ:LEMME-MELLIN1}
\een
where
\[
U(z) = \sum_{\ell \in {\qZ}\setminus \{0\}}\Gamma(\chi_{\ell}) e^{-2i\ell \pi z},\ %
\quad \mbox{with}\ \chi_{\ell}\equiv \frac{2 i \ell \pi}{\log{2}}\,.
\]
The Fourier series $U(z)$ has mean value $0$ and the amplitude of the
series does not exceed $10^{-6}$. ($\Gamma(z)$ is the Euler function
$\Gamma(z) =  \int_{0}^{\infty} e^{-t} t^{z-1} dz$.)
\end{lem}

\begin{proof} Asymptotics on the finite sum in equation~(\ref{EQ:LEMME-MELLIN1})
is obtained by direct use of Mellin transform asymptotics~\cite{PHILIPPE-ROBERT}.
Periodic fluctuations are occurring under the form of the Fourier
series $U(\log_2{n})$. However, the Fourier coefficients of $U(z)$
decrease very fast, so that the amplitude of the Fourier series is
very tiny, viz. $|U(z)| \le 10^{-6}$ (see e.g.,~\cite{PHILIPPE-ROBERT}
or~\cite[p. 131]{KNUTH}).
Last, the error term $O(n/2^r)$ in~(\ref{EQ:LEMME-MELLIN1}) results
from the truncated summation ${\sum_{k=1}^{k=r} n/2^k e^{-n/2^k}}$.
\end{proof}

\begin{lem} \label{LEMME-MELLIN2}
Let ${r_1}\equiv {r_1}(n)$ and ${r_2}\equiv {r_2}(n)$, such that $r_i
\to \infty$, while $n/2^{r_2} \to 0$ and $n/2^{r_1} \to \infty$ when
$n \to \infty$. Then, for all positive integer $m$,
\ben
& & \sum_{k={r_1}}^{{r_2}} \l(\frac{n}{2^k}\r)^m %
\exp{\l(-\frac{nm}{2^k}\r)} \; \;=\;\; \frac{m!}{m^{m+1} \log{2}} \cr
& & \;+\; \frac{1}{m^m \log{2}}\,U_m\b(\log_2(n)\b)
\; + \; O\l(\frac{2^{{r_1}m}}{n^m}\r) %
\; + \; O\l(\frac{n^m}{2^{{r_2}m}}\r) %
\; + \; O\l(\frac{1}{n}\r), \label{EQ:LEMME-MELLIN2}
\een
with $\dis{\chi_{\ell}\equiv \frac{2 i \ell \pi}{\log{2}}}\,.$
For any $\xi \ge 0$ and any positive integer $m$, the above Fourier
series
\begin{equation} \label{EQ:UMZPHI}
U_m(z) = \sum_{\ell \in {\qZ}\setminus \{0\}} \Gamma(m-1+\chi_{\ell}) %
\exp{(- 2i \ell \pi z)}
\end{equation}
has mean value $0$ and the amplitude of the series does not exceed $10^{-5}$.
\end{lem}

\begin{proof}
Again, asymptotics on the summation in equation~(\ref{EQ:LEMME-MELLIN2})
is completed by using the Mellin transform and complex
asymptotics~\cite{PHILIPPE-ROBERT}.
The error terms $O(2^{r_1 m}/n^m)$ and $O(n^m/2^{r_2 m})$
in~(\ref{EQ:LEMME-MELLIN2}) also result from the ``doubly truncated''
summation: $r_1\le k\le r_2$.
\end{proof}

\pl
We also use the following
\begin{lem} \label{GEOMETRIC_INEQUALITIES}
Let $(X_i)_{i\geq 1}$ and $(Y_i)_{i\geq 1}$ be two sequences of
independent Bernoulli random variables, denoted $B(P_i)$ and
$B(Q_i)$, respectively, and such that $P_i\leq Q_i$ for any $i$. By
definition,
\[
\qP(X_i=1) = 1-\qP(X_i=0)=P_i\ \quad \textrm{and}\ \quad %
\qP(Y_i=1) = 1-\qP(Y_i=0)=Q_i.\]

Let $H = \inf\{j\, |\, X_j = 1\}$ and $K=\inf\{j\, |\, Y_j = 1\}$,
which may be regarded as a first success in each sequence $X_i$
and $Y_i$ (resp.). Then, the ``stochastic inequality'' $K\leq_S H$
holds. In other words, for any non-negative integer $k$,
\[
\qP(K\leq k)\geq \qP(H\leq k).
\]
Moreover, for any non-decreasing function $f$,
\begin{equation}\label{equ}
\qE(f(K))\leq \qE(f(H)).
\end{equation}
\end{lem}
The above Lemma is a standard result in probability theory.
It can be proven by constructing a probability space $\Omega$ in
which the sequences of r.v. $(X_i)$ and $(Y_i)$ ``live'': for every
$\omega$, $X_i(\omega)=1$ $\Rightarrow$ $Y_i(\omega)=1$. For any
$\omega\in \Omega$, $K(\omega)\leq H(\omega)$, and the stochastic
order is then a simple consequence of this ``sure'' order on $\Omega$.
Any nondecreasing function $f$ also satisfies
$f(K(\omega))\leq f(H(\omega)), \forall \omega\in \Omega$,
and~(\ref{equ}) holds.

\subsection{Analysis of Algorithm 1}
Assume that Algorithm 1 is in a given round $j$ and that $k$ satisfies
$1 \leq k \leq \lceil \alpha^{j} \rceil$. Let $p_j(n)$ be the probability
that one station is \textit{elected} in round $j$.
In that round, that is for $k$ ranging from $1$ to $\lceil \alpha^{j} \rceil$,
the stations decide to broadcast with the sequence of probabilities
$(1/{2^k})_{ 1 \leq k \leq \lceil \alpha^{j} \rceil}$. We have,
\ben
p_j & = & \sum_{k=1}^{\lceil \alpha^{j} \rceil} %
\frac{n}{2^k}\l(1-\frac{1}{2^k}\r)^{n-1}\; \times \; %
\prod_{{i=1 \atop i\neq k}}^{\lceil \alpha^{j}\rceil} %
\l(1 - \frac{n}{2^i}\l(1-\frac{1}{2^i}\r)^{n-1}\r) \cr
& & \cr
& = & \sum_{k=1}^{\lceil \alpha^{j} \rceil} %
\frac{n}{2^k}\l(1-\frac{1}{2^k}\r)^{n-1}\; \times \; %
\frac{1}{\l(1-\frac{n}{2^k}\l(1-\frac{1}{2^k}\r)^{n-1}\r)} %
\; \times\; \prod_{i=1}^{ \lceil \alpha^{j} \rceil } %
\l(1-  \frac{n}{2^i}\l(1-\frac{1}{2^i}\r)^{n-1} \r) \cr
& & \cr \cr
& = & \sum_{m=0}^{\infty} \sum_{k=1}^{\lceil \alpha^{j}\rceil} %
\l(\frac{n}{2^k} \, \l(1-\frac{1}{2^k}\r)^{n-1} \r)^{(m+1)} %
\;\times\; \underbrace{\prod_{i=1}^{\lceil \alpha^{j}\rceil } %
\l(1 - \frac{n}{2^i}\l(1-\frac{1}{2^i}\r)^{n-1}\r)}_{s_j(n)}. \label{GERARDO}
\een
\no
\begin{rem} \label{CRUCIAL}
Simple considerations show that when $2^{\alpha^{j}} \ll n$,
the probability $\b(1 - s_j(n)\b)$ to have an election in the $j$-th round
is almost $0$ for large $n$. This remark explains the occurrences of
the crucial values $n/{2^{\alpha^{j}}}$ and
$j^{\star} = \lceil \log_{\alpha}{\log_2{n}} \rceil$ in the analysis.
\end{rem}

The following Lemma~\ref{Lemma-BOUND-SJ} provides an upper
bound on
\ben
s_j(n) = \prod_{i=1}^{ \lceil \alpha^{j} \rceil } %
\l(1 - \frac{n}{2^i}\l(1-\frac{1}{2^i}\r)^{n-1} \r).
\een
\begin{lem} \label{Lemma-BOUND-SJ}
Let $j$ be increasing integers such that $j \ge j^{\star}(n)$, then
\[\limsup_n s_j(n) \le .1884.\]
\end{lem}

\begin{proof} For any given $i_1$, for all $i \geq i_1$,
\[
\l(1-\frac{n}{2^i}\B(1-\frac{1}{2^i}\B)^{n} \r) %
\leq \Bigg(1-\frac{n}{2^i} \exp{\l( -\frac{n}{2^i}\B(1+\frac{1}{2^i}\B)\r)} \Bigg) %
\leq  \l(1-\frac{n}{2^i}\exp{\l( -\frac{n}{2^i}\l(1+\frac{1}{2^{i_1}}\r)\r)} \r).
\]
Since $\alpha^{j}\to\infty$ and $n/{2^{\alpha^{j}}}\to 0$,
$\alpha^{j} \gg \log_2{n}$, and by choosing
$i_1 = \lceil \frac{1}{2}\log_2{n} \rceil $ we obtain
\be
s_j(n) & \leq & \prod_{i=i_1}^{\lceil  \alpha^{j}\rceil} \Bigg(1-\frac{n}{2^i} %
\exp{\l( -\frac{n}{2^i}\B(1+\frac{1}{2^{i_1}}\B) \r)} \Bigg) %
\le \exp{\l(-\sum_{m\geq 1} \frac{1}{m} \sum_{i=i_1}^{\lceil \alpha^{j}\rceil} %
\frac{n^m}{2^{im}} \exp{\l(-\frac{nm}{2^i}\B(1+\frac{1}{2^{i_1}}\B)\r)}\r)} \cr
& \le & .1883 \;+\; O\l(\frac{1}{\sqrt{n}}\r) \;+\; O\l(\frac{n}{2^{\alpha^{j(n)}}}\r).
\ee
The value $\exp{\l(- \sum_{m\geq 1} m!/(m^{m+2}\log{2})\r)} = .188209\ldots$
is numerically computed with Maple. The upper bound on $\limsup s_j(n)$
is derived by taking into account the fluctuations of the Fourier
series, up to $e^{10^{-5}}$ in our case, and the Lemma follows.
\end{proof}

Next, the following Lemma~\ref{0.3694} provides an upper bound on $p_j(n)$
(defined in~(\ref{GERARDO})).

\begin{lem} \label{0.3694}
Let $j$ be increasing integers such that $j \ge j^{\star}(n)$, then
\[\limsup_n p_j(n) \le .3694.\]
\end{lem}

\begin{proof} By Lemma \ref{LEMME-MELLIN2}, since $n/{2^{\alpha^{j}}} \to 0$,
we have
\ben
\sum_{k=1}^{ \lceil \alpha^{j} \rceil} \l(\frac{n}{2^k}\r)^{(m+1)} \, %
\exp{\l(-\frac{(m+1)n}{2^k}\r)} \;\sim \; %
\frac{(m+1)!}{(m+1)^{m+2} \log 2}\;+\; %
\frac{1}{(m + 1)^{m+1} \log{2}}\,U_m\b(\log_2(n)\b), \label{NELSON}
\een
where $U_m(z)$ is defined in~(\ref{EQ:UMZPHI}).
Summing on $m$ in equation~(\ref{GERARDO}) and using the techniques
in Lemma~\ref{Lemma-BOUND-SJ} yields the above upper bound on $\limsup p_j(n)$,
numerically computed with Maple.
\end{proof}

\pl \no \textbf{Proof of Theorem \ref{Theorem-ALGO1}:} %
Let $\j \equiv j^{\star}(n) = \lceil \log_{\alpha}\log_2
(n)\rceil$, which implies that $n/2^{\alpha^{(\j+1)}}\to 0$ when
$n \to \infty$. According to Lemma~\ref{0.3694}, if $n$ is large
enough, \ben \label{MARCEL}
1 - p_j\geq q_1\,\mathbb{I}_{j\geq j^{\star}+1},\ \quad %
\textrm{where}\ q_1 = 1 - .3695 = .6305.
\een

As a consequence, the number of rounds $n_1$ in Algorithm~1
is smaller (with respect to the stochastic order) than $n_1'=j^{\star} + G$,
where $G$ is a geometric r.v. with parameter $q_1$. Indeed, let
\[n_1 = \inf\{j \ | \ \mbox{the election occurs in round}\ j\}\]
and let the success probability in the $j$-th round be $P_j = 1 - s_j$
(the successes in different rounds being independent).
Then, $P_j\geq Q_j$, where $Q_j=q_1\,\mathbb{I}_{j\geq j^{\star}+1}$.
Taking $n'_1$ as the first success in a Bernoulli sequence with probability
$Q_j$, we obtain $n'_1$ as described above. Indeed, the first $j^{\star}$
trials fail, and afterwards, each trial results in a success with probability
$q_1$. The additive number of trials needed follows a geometric distribution
$G(q)$, and
\[\qE(n_1)\;\leq \;\qE(n_1')=j^{\star}+q_1^{-1}\;=\;\log_{\alpha}\log_2 (n)+O(1).\]
Let $T_1 \equiv T_1(n)$ be the time needed to elect a leader in Algorithm 1.
Since $n_1'$ is larger than $n_1$ for the stochastic order and
$r\mapsto \sum_{i=1}^{r} \lceil \alpha^i \rceil $ is non-decreasing,
by Lemma~\ref{GEOMETRIC_INEQUALITIES},
\ben
\qE(T_1) & = & \qE\l(\sum_{j=1}^{n_1}\lceil \alpha^j \rceil\r) %
\;\le\;\qE\l(\sum_{j=1}^{n_1'}\lceil \alpha^j \rceil\r)\; %
\le \; \sum_{k=1}^{+\infty}\sum_{j=1}^{j^{\star}+k}(1+\alpha^j) q_1(1-q_1)^{k-1}
\label{doublesum} \\
& \le & c_{q_1}(\alpha) \log_2 (n)\;+\; O(\log\log n). \label{ESPERANCE}
\een
Note that, during a round the mean number of awake times for a
given station is smaller than 1. Taking into account the large number
of rounds, the total number of awake time slots is shown to be
smaller than $2n\log_{\alpha}\log_2(n)\b(1+o(1)\b)$. Since
$\qP\b(n_1\le \j(1-\qe)\b)\to 0$ when $n\to \infty$, the above value is
asymptotically tight. \hfill $\square$

\begin{rem} It is easily seen that the algorithm and the convergence
of the double sum in~(\ref{doublesum}) (resp.) require conditions
$\alpha > 1$ and $\alpha (1-q_1) < 1$, with $1-q_1 = .3695$
(resp.). The value of $\alpha$ may thus be chosen in the range $1
< \alpha < 2.707\ldots$, so as to achieve a tradeoff between the
average execution time of the algorithm and the global awake time.
Thus, the minimum value of the constant $c_{q_1}(\alpha)$ is
$c_{q_1}(\widetilde{\alpha}) \simeq 8.837$, with
$\widetilde{\alpha} = 1.3361\ldots$
\end{rem}

\subsection{Analysis of Algorithm 2}

\no \textbf{Sketch of proof of theorem \ref{Theorem-ALGO2}.}
As already stated, two awake stations are needed in Algorithm~2:
the one is only sending and the other is listening (the witness). The
corresponding probability expresses along the same lines as
in~(\ref{GERARDO}) and, instead of $p_j(n)$, one has now in step $j$,
\ben \label{PAOLO}
p'_j(n)  & = & \sum_{k=1}^{\lceil \alpha^{j} \rceil}%
\frac{1}{2} \, \frac{{n \choose 2}}{4^k}\l(1-\frac{1}{2^k}\r)^{n-2}\, %
\frac{1}{\l(1-\frac{1}{2}\,\frac{{n \choose 2}}{4^k} %
\l(1-\frac{1}{2^k}\r)^{n-2}\r)}\, %
\prod_{i=1}^{ \lceil \alpha^{j} \rceil } %
\l(1-\frac{1}{2}\,\frac{{n \choose 2}}{4^i}\l(1-\frac{1}{2^i}\r)^{n-2}\r)\!.
\een
The computation is quite similar to the proof of Theorem~\ref{Theorem-ALGO1};
it uses technical Lemmas as shown in Subsection~\ref{TECHLEM}. Again,
asymptotics on $p'_j(n)$ in equation~(\ref{PAOLO}) is completed by use
of Mellin transform asymptotics. Periodic fluctuations also occur under
the form of a Fourier series, and after some algebra the Theorem follows.
In the case of Algorithm~2,
$\exp{\l(- \sum_{m\geq 1} m!/(2^m m^{m+2} \log{2})\r)} = .462\ldots$
(instead of $\exp{\l(- \sum_{m\geq 1} m!/(m^{m+2} \log{2})\r)} = .188\ldots$
in Algorithm~1). Then, computing $p'_j(n)$ leads to the sum
\[
\sum_{m>0} \sum_{k} \l(\frac{1}{2}\r)^{m} %
\l(\frac{{n \choose 2}}{4^k}\r)^{m} \l(1-\frac{1}{2^k} \r)^{(n-2)m} %
\;\sim \;\; \sum_{m>0} \frac{m!}{2^m m^{(m+1)} \log{2}} %
\;\sim \;.8274\ldots,
\]

Now, the  mean number of broadcasting stations is $n/2$ and the
mean number of witnesses in round $j$ is
\[\frac{1}{2}\,\sum_{k=1}^{\lceil \alpha^j \rceil}\frac{{n\choose 2}}{4^k} %
\l(1 - \frac{1}{2^k}\r)^{n-2} = \; O(1).\]
(Recall that a station becomes a witness iff it wakes up exactly when
there exists a single sender.)

Thus, the average number of awake time slots per station taking place
in a round equals 2 time slots (as in Algorithm~1) plus $1/2+O(1/n)$,
due to the awaking stations appearing in line~(9) of Algorithm~2.
Therefore, for any station, the expected
number of awake time slots is bounded from above by
$2.5 \log_{\alpha} \log_{2}(n)\,\b(1 + o(1)\b)$.

Note that with $q_2 = .6176\ldots$, $\alpha$ now
meets the condition $1 < \alpha < 2.61\ldots$; and the minimum value
of the constant $c_{q_2}(\alpha)$ is $c_{q_2}(\widetilde{\alpha}) \simeq 8.96$,
with $\widetilde{\alpha} = 1.3295\ldots$
\hfill $\square$

\begin{rem}
Algorithms~1 and 2 can be improved by starting from $k=k_0$, $k_0 >1$
in line~(3). Asymptotically, the running time of the algorithms remains
the same, but starting from $k=k_0$ reduces the awake
time slots, to $\b(1 + \epsilon\b)\log_{\alpha} \log_{2}(n)$ for Algorithm~1
and $\b(1.5 + \epsilon\b)\log_{\alpha} \log_{2}(n)$ for Algorithm~2,
respectively (with $\epsilon = 1/2^{k_0-1}$).
Yet, this makes the running time longer for small values of $n$.
Therefore, the knowledge of any lower bound on $n$ greatly helps.
\end{rem}

\section{Conclusion}
In this paper, we present two new randomized leader election protocols
in  $n$-station RN with no knowledge of $n$, under the
assumption of weak and strong no-CD RN, respectively.
The expected $O\b(\log(n)\b)$ time complexity of Algorithms~1
and 2 achieves a quasi-optimality (up to a constant factor), with
each station keeping awake for $O\b(\log\log(n)\b)$ time
slots in both algorithms.

Our main contribution is to propose a class of energy-efficient
and quasi-optimal leader election protocols for individual clusters
of an $n$-station RN. This class of double-loop algorithms uses a
parameter $\alpha$ which
serves for a time-tuner in adjusting  the tradeoff between the average
time complexity of algorithms and the awake time slots of the $n$ stations.
(The tradeoff is only obtained with respect to time upper bounds).
Next, our analyses provide upper bounds on the current variables.
%
Also, the algorithms presented and the analysis of their performance
improve on~\cite{OLARIUCONF}.
Such results pave the way to address the design and analysis of a
broad class of energy-efficient protocols in RN: e.g.~naming
protocols, emulation protocols of single/multi-hop radio
networks~\cite{Yehuda1}, respectively, etc.

\end{document}